\documentclass{appolb}
\usepackage{graphicx}

\begin{document}
\title{Competing of Sznajd and voter dynamics in the Watts-Strogatz network
\thanks{Presented at 25th Marian Smoluchowski Symposium on Statistical Physics, Cracow, Poland, 9-13 September, 2012}%
}
\author{Marcin Rybak, Krzysztof Ku{\l}akowski 
\address{
AGH University of Science and Technology, Faculty of Physics and Applied Computer Science, al. Mickiewicza 30, PL-30059 Krakow, Poland
}
\\
}
\maketitle
\begin{abstract}
We investigate the Watts-Strogatz network with the clustering coefficient $C$ dependent on the rewiring probability. The network is an area of two opposite contact processes,
where nodes can be in two states, S or D. One of the processes is governed by the Sznajd dynamics: if there are two connected nodes in D-state  all their neighbors 
become D with probability $p$. For the opposite process it is sufficient to have only one neighbor in state S; this transition occurs with probability 1. The concentration 
of S-nodes changes abruptly at given value of the probability $p$. The result is that for small $p$, in clusterized networks the activation of S nodes prevails.
This result is explained by a comparison of two limit cases: the Watts-Strogatz network without rewiring, where $C$=0.5, and the Bethe lattice where $C$=0.
\end{abstract}
\PACS{07.05.Tp; 64.60.aq; 64.60.ah}
  
\section{Introduction}

Our aim here is to study the results of competition of two types of contact processes (CP) \cite{cp} in the Watts-Strogatz network. The idea is as follows. The network nodes 
are assumed to be in one of two states, say S and D. One of CP is the voter dynamics \cite{lg}, where a randomly selected S node changes the state of one of his neighbor from D to S.
The other CP is the unifying step of the Sznajd dynamics \cite{sznajd,sz2}, where a randomly selected pair of neighboring nodes, once both D, change the state of all their neighbors 
from S to D.\\

In a recent text \cite{my}, a similar competition has been investigated on the Watts-Strogatz (WS) and the Erd\"os-R\'enyi (ER) networks. Also there, one process of activating nodes has 
been ruled by the voter dynamics, and the other was triggered by a pair of mutually connected nodes. The difference between that approach and the present work is that in paper \cite{my}
a node was activated by a pair of D nodes only if it was a nearest neighbor of both nodes of the pair. On the contrary, in the Sznajd dynamics used here all neighbors of each member 
of the D pair are activated \cite{sznajd}. We note that the third model of this family is the bootstrap percolation, where nodes of the activating pair are not necessary neighbors of each 
other \cite{adler,baxter}; the latter model is not investigated here. \\

Our motivation is as follows. In \cite{my}, the clustering coefficient $C$ was used as a control parameter. In WS, $C$ was controlled by rewiring, while in ER, it was controlled by
adding new links, along the method proposed by Holme and Kim \cite{hol,amk}. The results of \cite{my} indicate, that in WS, tuning of $C$ can lead to switching the D process off and on.
Namely, the D process is blocked if the clustering coefficient is below some critical value. This effect is a consequence of the model assumption, that both nodes of the 
activating pair must be neighbors of the activated node. The density of triangles is controlled by $C$. If the triangles in the network are rare, the D process is stopped. To check this 
conclusion, we are going to investigate the same process with the Sznajd dynamics, where triangles are not necessary. Consequently, the clustering coefficient $C$ should be less 
important. \\

It is worth to add that the competing CP's enables a more precise measurement of intensity of the investigated process by its comparison with the voter dynamics. In a finite network,
in particular a small-world network, the process can reach the whole lattice in a few steps. To switch another process on allows to find their mutual intensity when both processes
balance each other. In \cite{my}, the voter process was more intense and therefore it was applied with probability $p$, while the D process was applied with probability 1. Here the 
situation is opposite; the Sznajd dynamics is applied with the probability $p$, and the voter dynamics with the probability 1. We look for the values of $p_c$ where the stationary 
percentage of both kinds of nodes is 50-50.\\

\section{Algorithm}

Fifty WS networks of $N$=1000 nodes and degree $k=4$ are prepared by gradual rewiring of randomly selected links, and they are stored when their clustering coefficient $C$
is in the range $(C_k,C_k+0.01)$. The demanded $C_k$ are from 0.5 down to 0.1. Next we select randomly a given 
percentage of nodes and assign them to be in the state S; other nodes are in the state D.\\

The simulation is performed as follows. A node is selected randomly; if it is in the state S, we check all its neighbors. If there is a node D, it is converted to be S. If the randomly 
selected node is in the state D, we continue with the probability $p$: namely, we check all his neighbors. If there is a node D, all its neighbors and all the neighbors of the initially 
selected node (which appeared to be D) are converted to be D. Having done this, we select another node and continue. The change is done immediately.\\

In each time step, $N$ nodes are selected. There is 2500 time steps. The results are averaged over 50 networks, except in the case $C=0.5$, when there is only one network and the
simulation is run 50 times.\\

\section{Results}

The influence of the clustering coefficient $C$ on the population of S nodes is most apparent in the range of small initial population of these nodes, then the results to be shown here
are obtained in this range. Also, their dependence on the initial state is visible there. \\
 
In Figs. 1 and 2 we show the time dependence of the population of S nodes for two values of the clustering coefficient $C$, 0.1 and 0.5. We observe that for $C$=0.1 (Fig.1) 
the time of simulation is long enough to get the stationary state. Although we have not obtained this state for $C$=0.5 (Fig. 2), the results clearly show that in the latter case 
the relatively large probability $p$ of the D process is not enough to reduce the population of S nodes as strongly, as it is found for $C$=0.1. Namely, while $p$=0.27 is enough to 
eliminate S nodes for $C$=0.1, we find that for $C$=0.5 S nodes still prevail for $p$=0.38.\\

The results on the population of S nodes after 2500 time steps are shown in Figs. 3 and 4 for different values of $C$ and two different values of the initial population of S nodes,
0.1 and 0.25. The results show how the fall of the observed population of S nodes with the probability $p$ depend on the clustering coefficient $C$. Clearly, the larger $C$, the 
stronger is the process S, as larger values of $p$ are necessary to damp the population of S nodes.\\

From these results, we can evaluate the probability $p_c$ when the calculated population of S nodes is 0.5 after 2500 time steps. As shown in Fig. 5, the initial population of S nodes does not change 
these results qualitatively. In both cases, the observed probability $p_c$ increases with $C$ from about 0.26 to about 0.4 in the investigated range of $C$.\\

\begin{figure}[htb]
\centerline{%
\includegraphics[width=12.5cm]{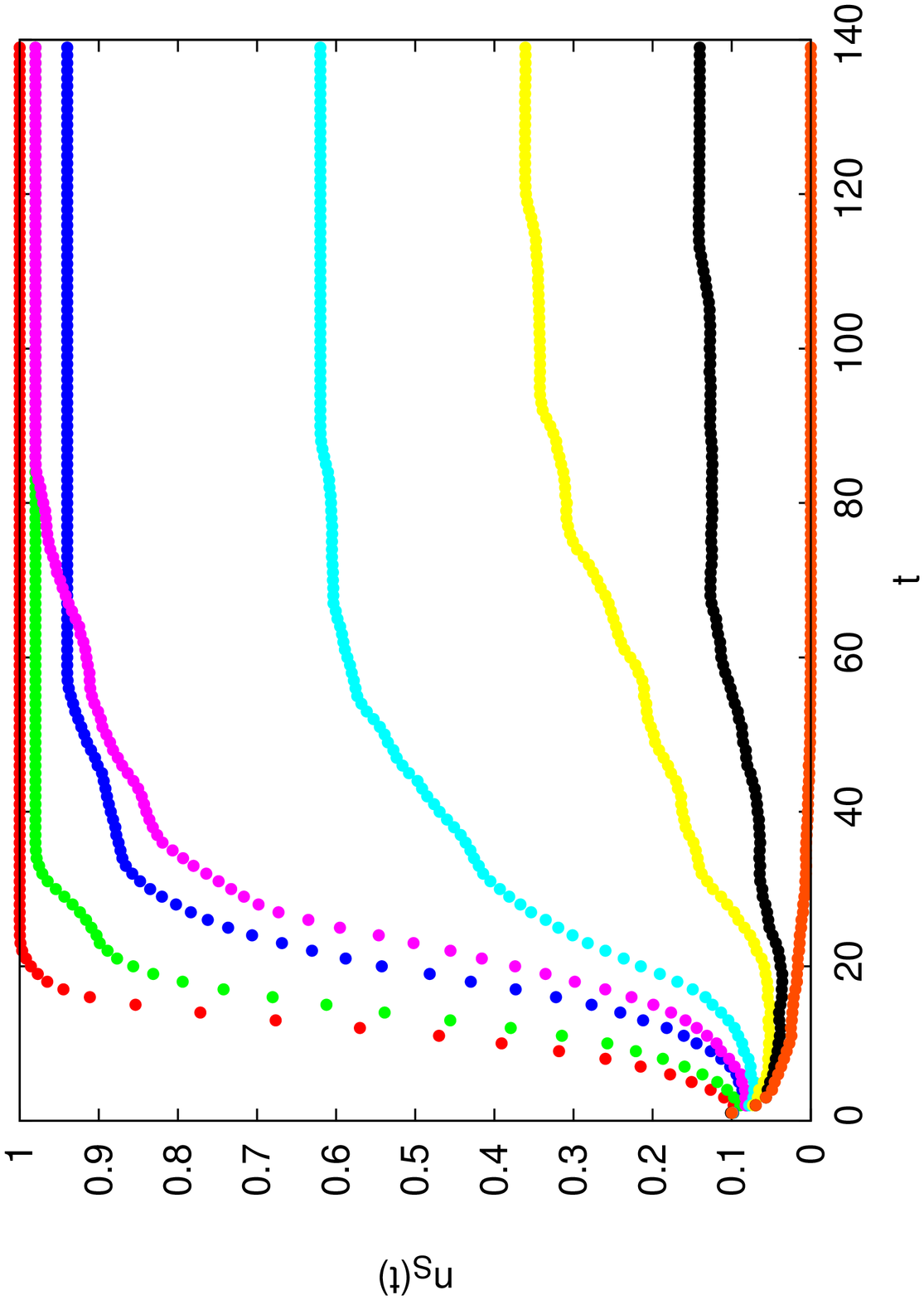}}
\caption{Time dependence of the S-nodes participation in the WS network (10 percent of all 1000 nodes are initially S-nodes) with $C$=0.1. 
The probability $p$ is from 0.2 to 0.27 with step 0.01, in ascending order from the top to the bottom of the graph.}
\label{Fig1}
\end{figure}

\begin{figure}[htb]
\centerline{%
\includegraphics[width=12.5cm]{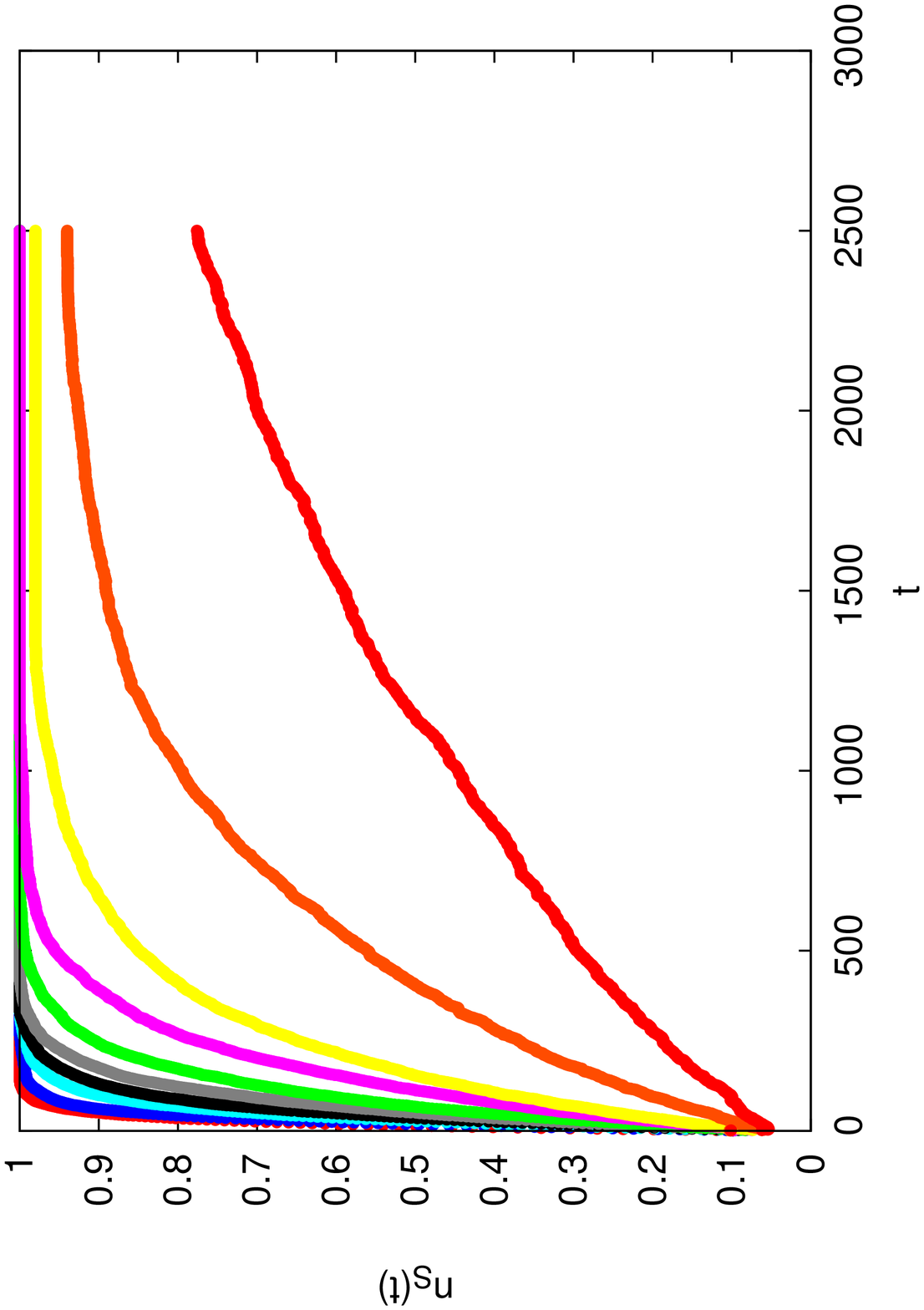}}
\caption{Time dependence of the S-nodes participation in the WS network (10 percent of all 1000 nodes are intially S-nodes) with $C$=0.5. 
The probability $p$ is from 0.20 to 0.38 with step 0.02, in ascending order from the top to the bottom of the graph.}
\label{Fig2}
\end{figure}

\begin{figure}[htb]
\centerline{%
\includegraphics[width=12.5cm]{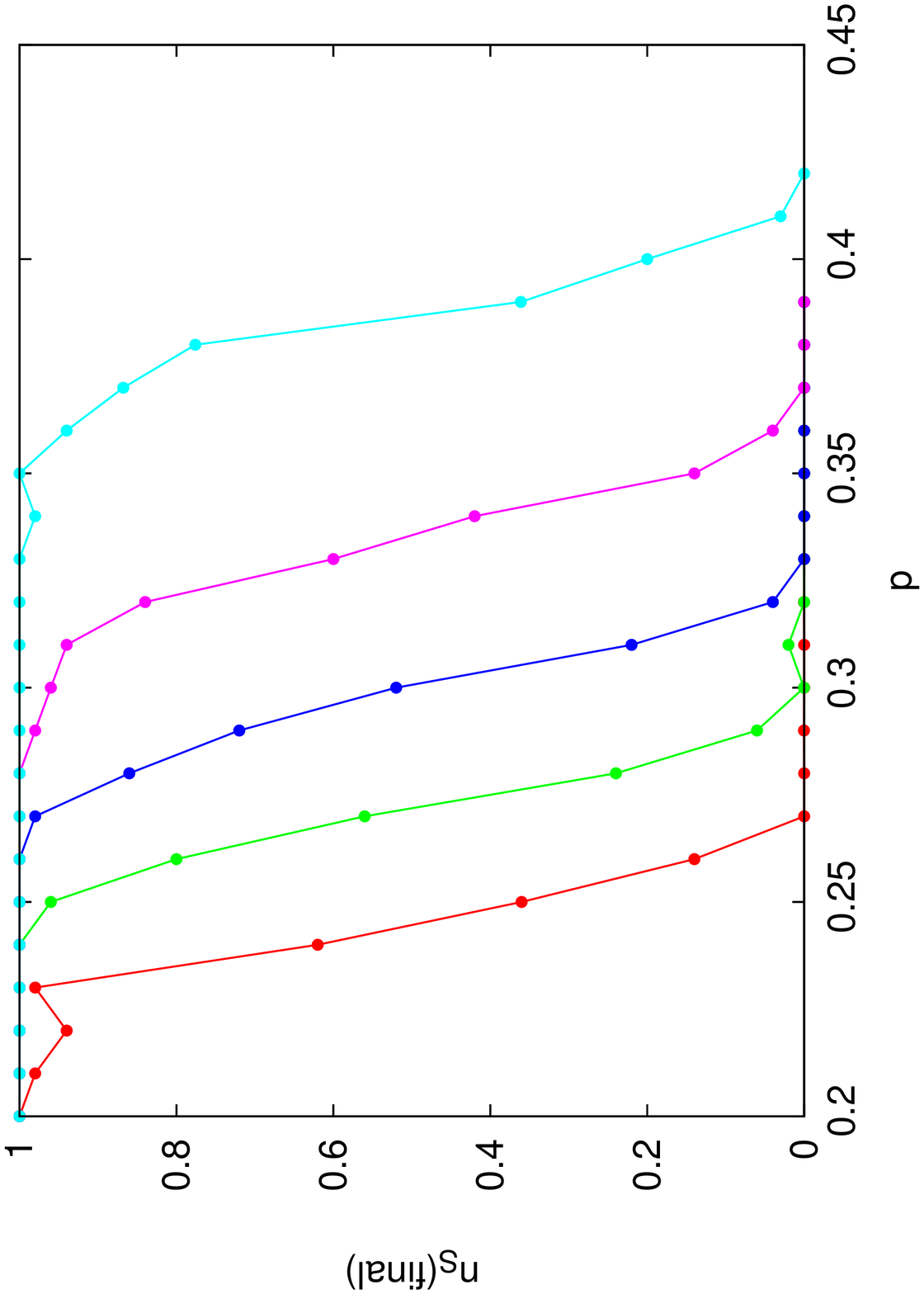}}
\caption{The final S-nodes participation vs the probability $p$ for
the WS network (10 percent of all 1000 nodes are initially S-nodes) for
different clustering coefficient $C$= 0.1, 0.2 , 0.3, 0.4 and 0.5 (curves from left to
right, respectively)}
\label{Fig3}
\end{figure}

\begin{figure}[htb]
\centerline{%
\includegraphics[width=12.5cm]{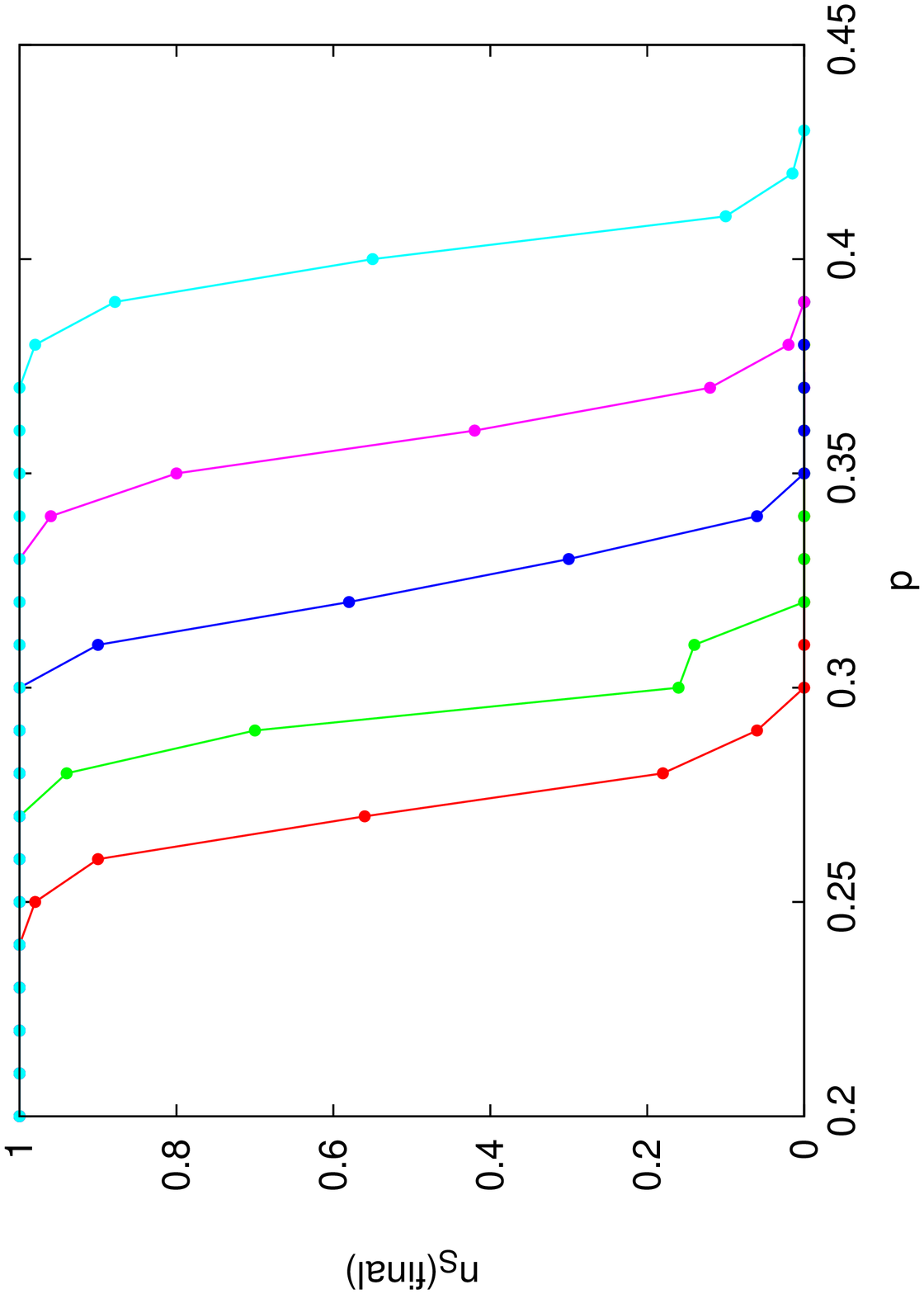}}
\caption{The final S-nodes participation vs the probability $p$ for
the WS network (25 percent of all 1000 nodes are initially S-nodes) for
different clustering coefficient $C$= 0.1, 0.2 , 0.3, 0.4 and 0.5 (curves from left to
right, respectively)}
\label{Fig4}
\end{figure}

\begin{figure}[htb]
\centerline{%
\includegraphics[width=12.5cm]{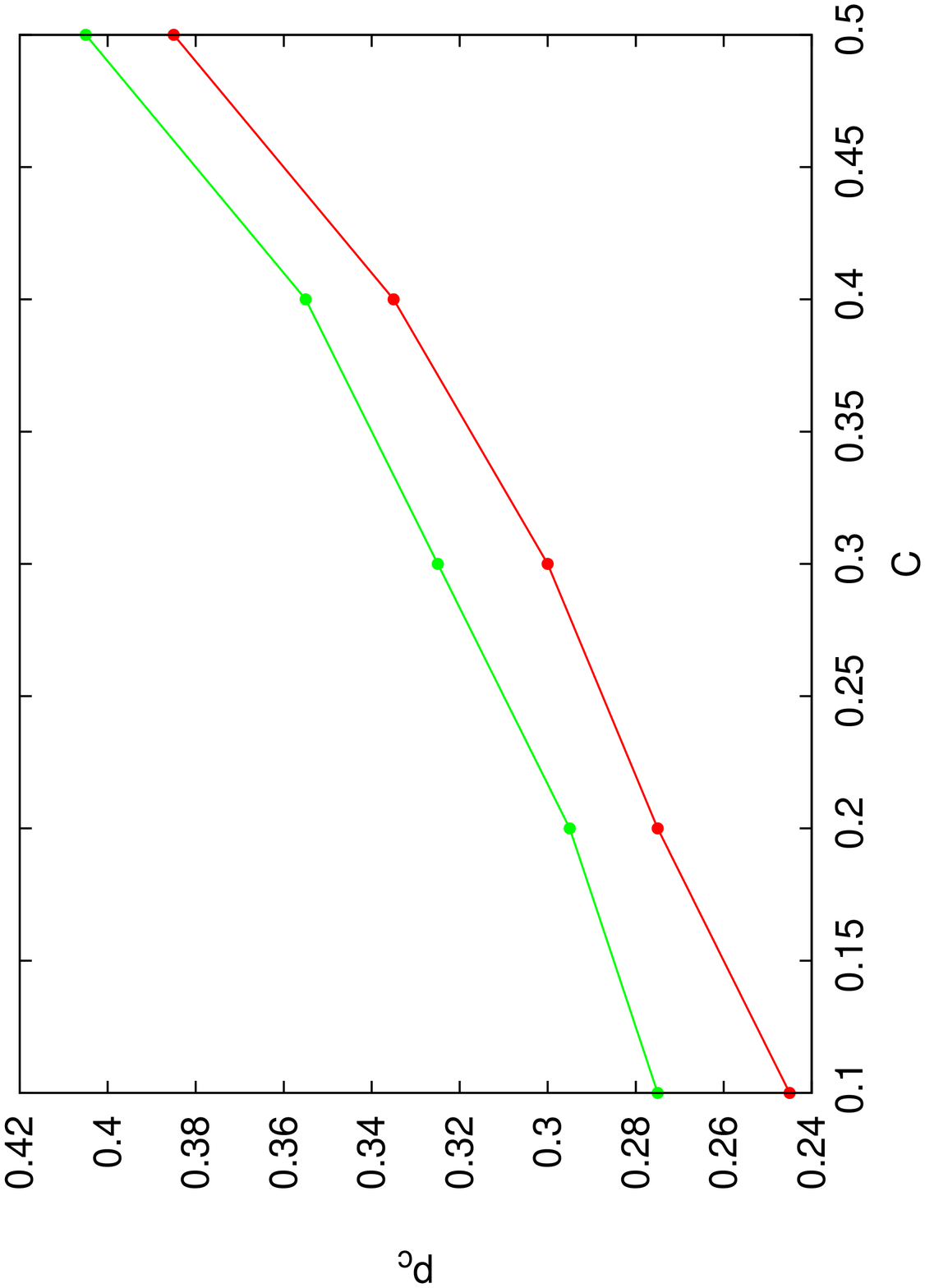}}
\caption{The critical probability $p_c$ against the clustering coefficient $C$ for
the WS network for different initial S-nodes participations equal to
10 percent (bottom curve) and 25 percent (top curve).}
\label{Fig5}
\end{figure}

\section{Discussion}

The results indicate, that the S process is relatively more active for WS networks with large clustering coefficient $C$. In particular, for the case when the initial 
amount of S nodes is 0.1, the critical value $p_c$ is about 0.38 for $C=0.5$, but only 0.25 for $C=0.1$. When the initial amount of nodes is 0.25, the same numbers for $p_c$
are 0.40 and 0.28. This means, that in clusterized networks, a larger probability of the D process is necessary to balance the S process, than for the case when the clustering
is small. \\

This result is opposite to the one obtained in our previous calculations \cite{my}, where the activation of D nodes was possible only if a node was simultaneously a 
nearest neighbor of both nodes of the activating pair of D nodes. This detail appears to be the cause of the observed different role of the clustering coefficient. While 
it seems natural that in our case the variation of $C$ is simply not relevant, the reversed influence of the clustering calls for an interpretation. \\

To accomplish this, let us compare two limit cases: the WS network without rewiring, where the clustering coefficient is maximal, and the Bethe lattice. For the mean degree $<k>$=4,
as assumed here, the respective values of $C$ are 0.5 and 0.0. Although in our case the WS network with large amount of rewiring is not equivalent to the Bethe lattice,
it is anyway close to an Erd\"os-R\'enyi network, which could be approximated by a regular tree. For the purpose of our explanation, it is important only that in both cases 
the clustering coefficient is small. Let us consider the simplified case of the velocity of spread of the D phase from a pair of neighbours, with the S process switched off 
and all nodes are simultaneously updated. In the WS network, this velocity is constant: for $k$=4, two nodes are switched from S to D at each step. On the contrary, in the Bethe 
lattice with the same degree the number of nodes switched from S to D along one tree branch is multiplied by $k-1$ at each time step. We should add that for the process 
investigated in \cite{my}, the above multiplication does not occur; in trees the D process considered there does not work at all.\\

The list of applications of the Sznajd model in politics and economy can be found in \cite{sz2}. Here we add that the comparison of the results presented here and of those in \cite{my}
reveals that a subtle modification of the mechanism seriously alters the intensity of the Sznajd dynamics in clusterized networks. In the literature on the Sznajd model, often we find a 
reference to the threshold effect introduced to sociology by Granovetter \cite{gran,mass}. In this effect, people are inclined to imitate the others' behaviour if more than a given
number of persons behave in a given way. This means, that the imitators see all the imitated persons, and not only one. Accordingly, this sociological effect is described more accurately
by the model described in \cite{my} or even by the bootstrap percolation, than by the Sznajd dynamics.\\

\section*{Acknowledgements} The calculations were performed in the ACK Cyfronet, Cracow, grants No. MNiSW/ IBM\_BC\_HS21/AGH/070/2010 and MNiSW/SGI3700/ AGH/070/2010. This work was 
partially supported from the AGH UST project No. 10.10.220.01.

\end{document}